\documentclass[prl,twocolumn,showpacs,a4paper]{revtex4}
\usepackage{epsfig}
\usepackage{graphicx}
\usepackage{mathrsfs,amsmath,amssymb}
\def\sH{{\mathcal H}}
\def\map#1{{\mathrm{#1}}}
\def\kk{\rangle\!\rangle}\def\bb{\langle\!\langle}
\def\Tr{\operatorname{Tr}}
\def\transp#1{{#1}^\tau}
\def\SU2{\mathbb{SU}(2)}
\def\<{\langle}\def\>{\rangle}
\def\geq{\geqslant}\def\leq{\leqslant}
\begin{document}
\title{Superbroadcasting of mixed states}

\author{Giacomo Mauro D'Ariano}\email{dariano@unipv.it}
\affiliation{{\em QUIT Group} of the INFM, unit\`a di Pavia}
\homepage{http://www.qubit.it} \affiliation{Dipartimento di Fisica
  ``A. Volta'', via Bassi 6, I-27100 Pavia, Italy}  
\author{Chiara Macchiavello}\email{chiara@unipv.it} 
\affiliation{{\em QUIT Group} of the INFM, unit\`a di Pavia}
\homepage{http://www.qubit.it} \affiliation{Dipartimento di Fisica
  ``A. Volta'', via Bassi 6, I-27100 Pavia, Italy} 
\author{Paolo Perinotti}\email{perinotti@fisicavolta.unipv.it} 
\affiliation{{\em QUIT Group} of the INFM, unit\`a di Pavia}
\homepage{http://www.qubit.it} \affiliation{Dipartimento di Fisica
  ``A. Volta'', via Bassi 6, I-27100 Pavia, Italy} 
\date{\today}

\begin{abstract}
  We derive the optimal universal broadcasting for mixed states of qubits.
  We show that the no-broadcasting theorem cannot be generalized to more than a single input copy.
  Moreover, for four or more input copies it is even possible to purify the input states
  while broadcasting. We name such purifying broadcasting {\em superbroadcasting}.
\end{abstract}
\pacs{03.65.-w, 03.67.-a}\maketitle
{\em Broadcasting}---namely distributing information over many users---suffers  in-principle
limitations when the information is quantum, and this poses a critical issue in quantum information
theory, for distributed processing and networked communications. For pure states an ideal broadcasting
coincides with the so-called {\em quantum cloning}, corresponding to an ideal device capable of
producing from a finite number $N$ of copies of the same state $|\psi\>$ a larger number $M>N$ of
output copies of the same state, for a given set of input states.  
Since such a transformation is not
isometric, it cannot be achieved by any physical machine on a generally nonorthogonal set of states
(this is essentially  the content of the {\em no-cloning} theorem \cite{Wootters82,Dieks82,Yuen}). The 
situation is more involved when the states are mixed, since from the point of view of each single
user the local mixed 
state is indistinguishable from the partial trace of an entangled state, and there are infinitely
many joint states corresponding to ideal broadcasting. For this reason in the literature
\cite{fuchs} the word {\em broadcasting} is used technically to denote a map whose output has
identical local states, versus the word {\em cloning} used for the case of tensor product of identical states.
\par Since ideal cloning is not possible, the quantum information encoded on pure states can be
broadcast only approximately, and this posed the problem of optimizing the broadcasting e. g. by
maximizing an input-output fidelity equally well on all pure states. In the literature this kind of optimized 
broadcasting is called {\em optimal universal cloning} \cite{Buzek,Gisin,sdc,Werner}.  For mixed states
the no-cloning theorem is not logically sufficient to forbid ideal 
broadcasting.  In Ref. \cite{fuchs} the impossibility of broadcasting
has been proved  in the case of one input copy and two output copies for a set 
of density operators generally non mutually commuting. Later, in the literature (see, for example,
Ref. \cite{clifton}) this result has been often implicitly considered as the generalization of the
no-cloning theorem to the case of mixed input states. In the present paper we
will show that this assertion cannot be generalized to more than a single input copy. In particular,
for numbers of input copies $N\geq 4$ the no-broadcasting theorem does not hold, and it is even
possible to purify while broadcasting. We  
named such a procedure  {\em superbroadcasting}.
\begin{figure}[h]
\epsfig{file=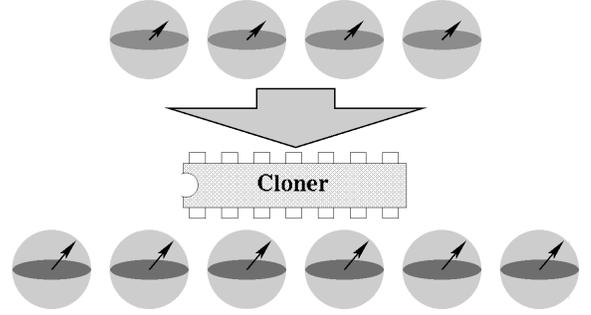,width=8cm}
\caption{With four or more input copies the no-broadcasting theorem can be violated. One can actually 
increase the purity of local states while broadcasting, corresponding to a stretching of the Bloch vector.  In this
  purifying broadcasting mechanism, called {\em superbroadcasting}, the available information on the
  state of the input copies cannot increase due to the detrimental correlations among the output copies.}
\end{figure}
We now present the theoretical derivation of our result.
\par Let us consider a general broadcasting channel from $N$ to $M$ copies,
namely a completely positive (CP) trace-preserving map from states on
$\sH_\mathrm{in}\doteq\sH^{\otimes N}$ to states on
$\sH_\mathrm{out}\doteq\sH^{\otimes M}$ that is invariant under
permutations of input copies and of output copies. Moreover, we take
the broadcasting to be universal, namely the broadcasting map $\map{B}$
is covariant under
the group of unitary transformations of $\sH$, more precisely
\begin{equation}
\map{B}(U^{\otimes N}\rho^{\otimes N}U^\dag{}^{\otimes N})=
U^{\otimes M}\map{B}(\rho^{\otimes N})U^\dag{}^{\otimes M}.\label{covariance}
\end{equation}
We will restrict attention to qubits, namely $\sH\simeq{\mathbb C}^2$. Upon
using the Choi-Jamiolkowsky representation \cite{dalop}
\begin{equation}
\begin{split}
  R_\map{B}=&\map{B}\otimes\map{I}(|I\kk\bb I|),\\ \map{B}(Q)=&
  \Tr_\mathrm{in}[(I_\mathrm{out}\otimes\transp{Q})R_\map{B}]
\end{split}
\end{equation}
where $Q$ denotes a state on $\sH_{in}$, and $R_\map{B}$ is a positive operator on
$\sH_\mathrm{out}\otimes\sH_\mathrm{in}$, the
covariance condition (\ref{covariance}) is equivalent to invariance of
$R_\map{B}$ under the group representation $U_g^{\otimes M}\otimes
U_g^*{}^{\otimes N}$, $U_g$ denoting the $j=\frac{1}{2}$ representation, for $g\in\SU2$  [the symbol
$|I\kk$ denotes the maximally entangled vector $|I\kk=\sum_{n}|n\>\otimes|n\>$, and $\transp{}$§
denotes transposition with respect to the orthonormal basis $\{|n\>\}$]. In the Choi-Jamiolkowsky
representation the trace-preserving condition on the CP map reads
\begin{equation}
\Tr_\mathrm{out}[R_\map{B}]=I_\mathrm{in}\,,
\end{equation}
where $I_\mathrm{in}$ denotes the identity on $\sH_\mathrm{in}$. For
the unitary group $\SU2$ the complex conjugate representation of any
unitary representation, say $V_g$, is unitarily equivalent to the
direct representation, i.~e.  $V_g^* =CV_gC^\dag$, under the $\pi$-rotation $C$ around the $y$ axis. The 
explicit form of $C$ actually depends on the particular representation $V_g$: for the tensor
representation $U_g^{\otimes N}$ one has $C\equiv  i\sigma_y^{\otimes N}$. It is then convenient to
rewrite the map as follows 
\begin{equation}
\map{B}(Q)=\Tr_\mathrm{in}[(I_\mathrm{out}\otimes\tilde{Q})S_\map{B}]
\label{eq:mappa}
\end{equation}
with
\begin{equation}
\tilde{Q}\doteq C\transp{Q}C^\dag,
\qquad S_\map{B}\doteq (I_\mathrm{out}\otimes C)R_\map{B}(I_\mathrm{out}\otimes C^\dag),
\end{equation}
and now covariance of the CP map $\map{B}$ corresponds to invariance
of $S_\map{B}$ under the representation $U_g^{\otimes (N+M)}$. A
tensor product representation $U_g^{\otimes L}$ decomposes into
irreducible components according to the Wedderburn decomposition of
spaces
\begin{equation}
\sH^{\otimes L}=\bigoplus_{j=\bb L/2\kk}^{L/2}\sH_j\otimes{\mathbb C}^{d_j},\label{CB}
\end{equation}
where $\bb x\kk$ denotes the fractional part of $x$ (i. e. $\bb
L/2\kk=0$ for $L$ even and $\bb L/2\kk=1/2$ for $L$ odd), and the multiplicity $d_j$ can be
evaluated by recurrence on $L$ by adding a qubit at a time, giving 
$d_j=\frac{2j+1}{L/2+j+1}\binom{L}{L/2+j}$ \cite{cirekma}. Eq. (\ref{CB}) is also called {\em
Clebsch-Gordan series}. The spaces $\sH_j$ and ${\mathbb C}^{d_j}$ 
are called {\em representation} and {\em multiplicity} spaces,
respectively. With the above decomposition the group representation
writes $U_g^{\otimes L}=\oplus_{j=\bb L/2\kk}^{L/2} U_g^{(j)}\otimes
I_{d_j}$, whereas an operator invariant under $U_g^{\otimes L}$ has
the form $\oplus_{j=\bb L/2\kk}^{L/2} I_j\otimes W^{(j)}$, $I_j$
denoting the identity over the representation space $\sH_j$, and
$W^{(j)}$ an operator on the multiplicity space ${\mathbb C}^{d_j}$. On the
other hand, an operator invariant under the permutation group
$\mathbb{P}_L$ of the $L$ copies of the representation has the form
$\oplus_{j=\bb L/2\kk}^{L/2} Z_j\otimes I_{d_j}$, where $Z_j$ is any operator
on the representation space $\sH_j$ (this is the so-called Schur-Weyl
duality) \cite{ford}. Since the operator $S_\map{B}$ is invariant
under $\mathbb{P}_M\times\mathbb{P}_N$ it must be of the form
$S_\map{B}=\oplus_{j=\bb M/2\kk}^{M/2}\oplus_{l=\bb
  N/2\kk}^{N/2}S_{jl}\otimes I_{d_j}\otimes I_{d_l}$, where $S_{jl}$
is a positive operator over $\sH_j\otimes\sH_l$. By further
decomposing $\sH_j\otimes\sH_l=\oplus_{J=|j-l|}^{j+l}\sH_J$ into
invariant subspaces and imposing invariance of $S_\map{B}$ under
$U_g^{\otimes (M+N)}$, one obtains the general form
\begin{equation}
S_\map M=\bigoplus_{j=\bb M/2\kk}^{M/2}\bigoplus_{l=\bb N/2\kk}^{N/2}
\bigoplus_{J=|j-l|}^{j+l} s_{j,l,J}P_{J}^{(j,l)}\otimes{I_{d_j}}\otimes{I_{d_l}}\,,
\end{equation}
for positive coefficients $ s_{j,l,J}$, $P_{J}^{(j,l)}$ denoting the
orthogonal projector over the irreducible representation $J$ coming
from the couple $j,l$.\par

The trace preservation condition is now equivalent to
\begin{align}
&\Tr_\mathrm{out}[S_\map M]=\\ &\sum_{j=\bb
    M/2\kk}^{M/2}\bigoplus_{l=\bb N/2\kk}^{\frac
    N2}\Tr_j\left[\bigoplus_{J=|j-l|}^{j+l} d_j
    s_{j,l,J}P_J^{(j,l)}\right]\otimes{I_{d_l}}=I_\mathrm{in}\,.\nonumber
\end{align}
Since $\Tr_j[P_J^{(j,l)}]$ is invariant under $U_g^{(l)}$, one can
easily see that $\Tr_j[P_J^{(j,l)}]=\frac{2J+1}{2l+1}I_l$, whence the
latter condition becomes
\begin{equation}
\bigoplus_{l=\bb N/2\kk}^{N/2}\sum_{j=\bb M/2\kk}^{M/2}
\sum_{J=|j-l|}^{j+l} d_j s_{j,l,J}\frac{2J+1}{2l+1}I_l\otimes{I_{d_l}}=I_\mathrm{in}\,,
\end{equation}
namely
\begin{equation}
\sum_{j=\bb M/2\kk}^{M/2}\sum_{J=|j-l|}^{j+l} d_j s_{j,l,J}\frac{2J+1}{2l+1}=1\,,\quad\forall 
\bb N/2\kk\leq l\leq\frac N2\,,
\label{eq:tracepres}
\end{equation}
with positive coefficients $s_{j,l,J}$.\par 

Upon writing the input state $\tilde{Q}=\tilde{\rho}^{\otimes N}$ in
the Bloch vector form, we have the decomposition
\begin{equation}
\begin{split}
&\tilde{\rho}^{\otimes N}=\left[\tfrac{1}{2}(I-r\vec k\cdot\vec\sigma)\right]^{\otimes N} \\
&=(r_+r_-)^{N/2}
\bigoplus_{l=\bb N/2\kk}^{N/2}\sum_{n=-l}^l\left(\frac{r_-}{r_+}\right)^n|ln\>\<ln|\otimes I_{d_l}\,,
\end{split}
\label{eq:decomp}
\end{equation}
where $0\leq r\leq 1$, and $r_\pm\doteq \frac{1}{2}(1\pm r)$, and
$|ln\>$ denotes the eigenstate of the angular momentum component $\vec
k\cdot\vec J ^{(l)}$ with eigenvalue $n$. From Eq.
(\ref{eq:tracepres}) we see that the broadcasting channels from $N$ to
$M$ make a convex set, with the extreme points classified by functions
$\varphi$ and $\Phi$ corresponding to a given choice $j=\varphi(l)$,
$J=\Phi(l)$, namely to the choice of coefficients
\begin{equation}
s_{j,l,J}^{(\varphi,\Phi)}=\frac{2l+1}{2J+1}\frac 1{d_j}\delta_{j,\varphi(l)}\delta_{J,\Phi(l)}\,,
\end{equation}
or to the Choi-Jamiolkowsky operator
\begin{equation}
S_\map M^{(\varphi,\Phi)}=\bigoplus_{l=\bb N/2\kk}^{N/2}\frac{2l+1}{2\Phi(l)+1}\frac 1{d_{\varphi(l)}}P^{(\varphi(l),l)}_{\Phi(l)}\otimes{I_{d_{\varphi(l)}}}\otimes{I_{d_l}}\,.
\label{eq:extmap}
\end{equation}
Using the expression \eqref{eq:extmap} for extremal broadcasting
channels and Eq. \eqref{eq:decomp} for the input state we can evaluate
the output state
\begin{equation}
\begin{split}
&\map{M}_{(\varphi,\Phi)}(\rho^{\otimes N})=(r_+r_-)^{N/2}
\bigoplus_{l=\bb N/2\kk}^{N/2}\frac{2l+1}{2\Phi(l)+1}\frac1{d_{\varphi
(l)}}\\
&\times\sum_{n=-l}^l\left(\frac{r_-}{r_+}\right)^n\Tr_l[(I_{\varphi(l)}\otimes|ln\>\<ln|)P^{(\varphi(l),l)}_{\Phi(l)}]\otimes{I_{d_{\varphi(l)}}}\,.
\end{split}
\end{equation}
In terms of Clebsch-Gordan coefficients, this can be rewritten as
\begin{equation}
\begin{split}
&\map{M}_{(\varphi,\Phi)}(\rho^{\otimes N})=(r_+r_-)^{N/2}\\ &\times\sum_{l=\bb N/2\kk}^{N/2}\frac{2l+1}{2\Phi(l)+1}\frac{d_l}{d_{\varphi(l)}}\sum_{n=-l}^l\left(\frac{r_-}{r_+}\right)^n\\
&\times\!\!\!\sum_{m=-\varphi(l)}^{\varphi(l)}\<\Phi(l)m+n|\varphi(l)m,ln\>^2|\varphi(l)m\>\<\varphi(l)m|\otimes I_{d_{\varphi(l)}}.
\label{eq:output}
\end{split}
\end{equation}
Now, we are interested in the single output copy, which is the
broadcast state. This is given by the partial trace of Eq.
\eqref{eq:output} over $M-1$ copies. The evaluation of the partial trace needs the matching
between the Wedderburn decomposition and the qubit tensor product 
representation. According to the Schur-Weyl
duality the multiplicity space of the Wedderburn decomposition supports a unitary irreducible
representation of the permutation group $\mathbb{P}_M$ of the $M$ qubits. Therefore, one has the
identity for any operator $X_j$ on $\sH_j\otimes{\mathbb C}^{d_j}$
\begin{equation}
\sum_{l\in\mathbb{P}_M}\pi_l X_j\pi_l^\dag=\frac{M!}{d_j}\Tr_{{\mathbb C}^{d_j}}[X_j]\otimes I_{d_j}
\end{equation}
where $\pi_l$ denotes the generic permutation. In particular, for
$X_j=|jm\>\<jm|\otimes|1\>\< 1|$, $|1\>$ denoting any fixed
vector of ${\mathbb C}^{d_j}$, one has
\begin{equation}
|jm\>\<jm|\otimes I_{d_j}=
\frac{d_j}{M!}\sum_{l\in\mathbb{P}_M}\pi_l X_j \pi_l^\dag\label{permtrace}
\end{equation}
Clearly, one can always choose the given vector of the irreducible representation as
\cite{cirekma} 
\begin{equation}
|jm\>\otimes|1\>=|jm\>\otimes|\Psi_-\>^{\otimes \frac{M}{2}-j},
\end{equation}
where $|\Psi_-\>$ denotes the singlet. We can then take the partial trace of both sides of
Eq. (\ref{permtrace}). For each permutation, say $\pi_s$, which exchanges the last qubit with one
belonging to a singlet, one has $\Tr_{M-1}[\pi_s X_j \pi_s^\dag]=\frac{I}{2}$, and we have
$(M-2j)(M-1)!$ permutations of this kind. On the other hand, for each permutation, say $\pi_m$,
which exchanges the last qubit with one belonging to the $j$-multiplet, one has $\Tr_{M-1}[\pi_m X_j
\pi_m^\dag]=\Tr_{j-\frac12}[|jm\>\<jm|]$ and there are $2j(M-1)!$ permutations of this kind. Using the explicit form of the Clebsch-Gordan coefficients one 
can derive the following identity 
\begin{equation}
\Tr_{j-\frac12}[|jm\>\<jm|]=\frac{1}{2}I+\frac{m}{2j}\vec k\cdot\vec\sigma\,.
\end{equation}
Substituting the above formula when performing the partial trace of both sides
of Eq. (\ref{permtrace}), one obtains the following expression for the single
copy output density operator
\begin{equation}
\begin{split}
&\rho^\prime_{(\varphi,\Phi)}(r)=(r_+r_-)^{N/2}\sum_{l=\bb N/2\kk}^{N/2}\frac{2l+1}{2\Phi(l)+1}d_l\sum_{m=-\varphi(l)}^{\varphi(l)}\\
&\times\!\!\!\sum_{n=-l}^l\left(\frac{r_-}{r_+}\right)^n\<\Phi(l)m+n|\varphi(l)m,ln\>^2\,\frac12\left(I+\frac {2m}M\vec k\cdot\vec\sigma\right)\!.
\end{split}
\label{eq:singlecopy}
\end{equation}
We are now in position to analyse the broadcast state, in particular
its Bloch vector. In Eq. (\ref{eq:singlecopy}) we see that the input and the output Bloch vectors are
parallel, and clearly $[\rho^\prime,\rho]=0$. On the other hand, the length of the output Bloch
vector is given by 
\begin{equation}
\begin{split}
&r^\prime_{(\varphi,\Phi)}(r)=(r_+r_-)^{N/2}\sum_{l=\bb N/2\kk}^{N/2}\frac{2l+1}{2\Phi(l)+1}d_l\\
&\times\!\!\!\sum_{m=-\varphi(l)}^{\varphi(l)}\sum_{n=-l}^l\left(\frac{r_-}{r_+}\right)^n\<\Phi(l)m+n|\varphi(l)m,ln\>^2\,\frac {2m}M
\end{split}
\end{equation}
We are now interested in maximizing the length of the output Bloch vector. Since $r'$ is linear on
the convex set of broadcasting channels, we just need to consider extremal maps, and look for the
maximum $r'_{opt}(r)=\max_{(\varphi,\Phi)}\{ r^\prime_{(\varphi,\Phi)}(r)\}$. It is possible to
prove\cite{new} that the maximal $r^\prime_{(\varphi,\Phi)}(r)$ is achieved
for $\varphi(l)=M/2$ and for 
$\Phi(l)=\left|l-\frac{M}{2}\right|$, independently on $r$. For pure states these optimal maps
coincide with those of optimal universal cloning transformations\cite{Buzek,Gisin,sdc,Werner}. 
Also, it can be shown\cite{new} that our optimal map gives the same results achievable using the
procedure of Ref. \cite{cirekma}. 
\begin{figure}[h]
\epsfig{file=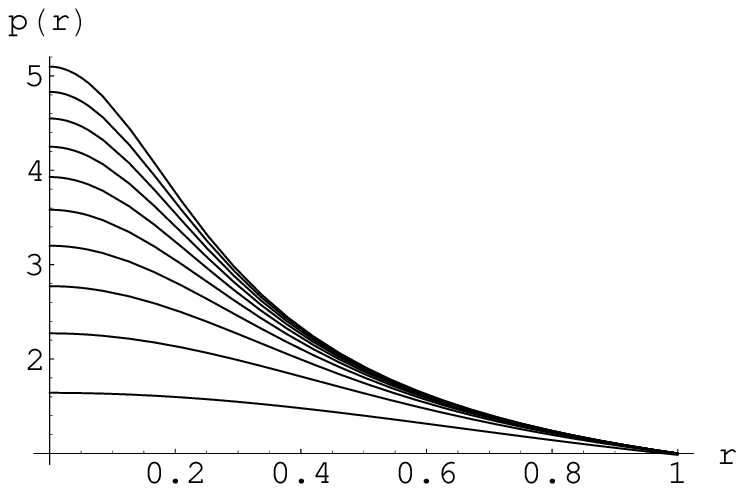,width=4.5cm}\epsfig{file=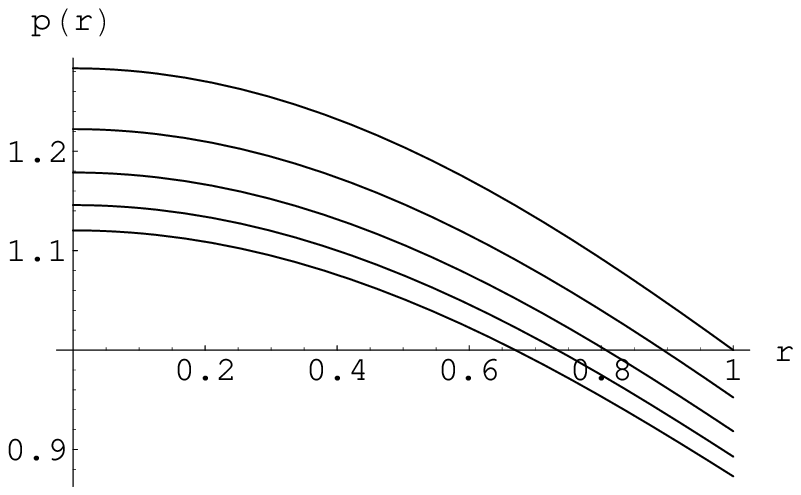,width=4cm}
\caption{The scaling factor $p(r)$ versus $r$. On the left: for $M=N+1$ and
  $N=10,20,30,40,50,60,70,80,90,100$ (from the bottom to the top. On the right:
for $N=5$ and $5\leq M\leq9$ (from the top to the bottom).\label{f:p(r)}}
\end{figure}
\par As an example, in Fig. \ref{f:p(r)} we plot the {\em scaling factor}
$p(r)=r'_{opt}(r)/r$ for the maps maximizing $r'$ for $N=5$ and several values of $M$. One can see that
for a wide range of values of $r$, one has $p(r)\geq 1$. This
corresponds to a purification of the local states, and since one also has a number of copies at the output 
$M> N$ greater than the number of inputs, it is actually a broadcasting with simultaneous purification,
what we call {\em superbroadcasting}. Clearly, for $M\leq N$ one has more purification than for $M>
N$, corresponding to the purification protocol \cite{cirekma}. The superbroadcasting occurs for
$N\geq 4$ input copies. As a rule, one has purification below some value $r_*(N,M)$ of the input purity, for
a bounded number $M\leq M_*(N)$ of the output copies. 
\begin{figure}[h]
\epsfig{file=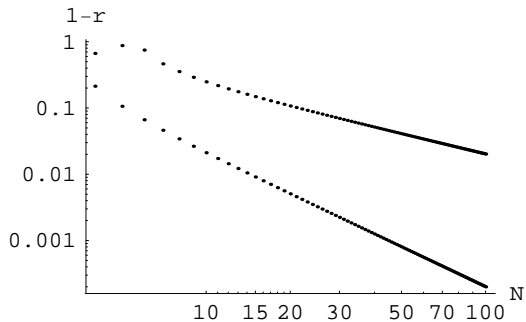,width=7cm}
\caption{Logarithmic plot versus $N$ of $1-r_*(N,N+1)$ (bottom) and $1-r_*(N,M_*(N)) $ (top), where
  $r_*(N,M)$ denotes the maximum purity for which one has superbroadcasting from $N$ to 
  $M$ copies, $M_*(N)$ being the maximum number of output copies for $N$ inputs (the area above the
  lower plot is the region in which superbroadcasting is possible). The two asymptotic behaviours
  are $N^{-1}$ and $2N^{-2}$. \label{f:rstar}} 
\end{figure}
In Fig. \ref{f:rstar} we plot $r_*(N,N+1)$  and $r_*(N,M_*(N))$  versus the number of input
copies $N$. After the threshold at 
$N=4$ corresponding to $r_*(4,5)=0.787$, one has a monotonic increase of $r_*(N,N+1)$ and
$r_*(N,M_*(N))$  toward asymptotic purity, with power laws $2 N^{-2}$ and $N^{-1}$,
respectively. For larger $M$ one has  
a generally higher threshold for $N$, and smaller values of $r_*(N,M)$. For $N=4$ one has
superbroadcasting for up to $M=7$, for $N=5$ up to $M=21$, and for $N=6$ up to $M=\infty$.
Notice that perfect broadcasting (corresponding to $p(r)=1$) can be achieved under the same conditions of
superbroadcasting, (clearly generally by a different map).
We remind that we have considered boradcasting of universally covariant sets of mixed
states. Indeed, for smaller sets of input states it can be shown that superbroadcasting is possible
also for $N=3$ input copies (as for equatorial phase-covariant mixed states\cite{new}), and, for
even smaller sets one cannot exclude superbroadcasting also for $N=2$. 
\par In conclusion, we have derived the optimal universal broadcasting for mixed states of qubits,
optimal in the sense that it maximizes the purity of local states. For pure states and $M>N$ the map
coincides with the optimal universal cloning transformation\cite{Buzek,Gisin,sdc,Werner}, whereas
for $N\geq M$ it is equivalent to the optimal purification map of Ref. \cite{cirekma}. Thus our optimal
broadcasting map generalizes/interpolates between optimal cloning and optimal purification. 
We have shown that the no-broadcasting theorem\cite{fuchs} for noncommuting mixed states cannot be
generalized to more than a single input copy,  and for $N\geq 4$ input copies one can even
purify the state while broadcasting, below some maximum value of the purity. We named such
phenomenon {\em superbroadcasting}. The possibility of 
superbroadcasting does not correspond to an increase of the available information about the original
input state $\rho$, due to detrimental correlations between the local broadcast copies, which does not allow
to exploit their statistics. This phenomenon was already noticed in Ref. \cite{keylwer}, in 
an asymptotic analysis of the rate of optimal purification procedures. Notice that the correlations alone
among qubits cannot be erased by any physical process, since the de-correlating map which sends a
state to the tensor product of its partial traces is non linear. From the point of view of
single users our broadcasting protocol is actually a purification (for states sufficiently mixed),
and the same broadcasting process transfers some noise from the local states to the correlations between
them.  We think that the present result opens new interesting perspectives in the ability of
distributing quantum information in a noisy environment.
\par This work has been co-founded by the EC under the program ATESIT (Contract No.
IST-2000-29681), and QUPRODIS (Contract No. IST-2002-38877). P.P. acknowledges support from the INFM
under project PRA-2002-CLON. G.M.D. acknowledges partial support by the MURI program administered by
the U.S. Army Research Office under Grant No. DAAD19-00-1-0177. 

\end{document}